\documentclass[fleqn,usenatbib]{mnras}

\usepackage{newtxtext,newtxmath,ulem}

\usepackage[T1]{fontenc}

\DeclareRobustCommand{\VAN}[3]{#2}
\let\VANthebibliography\thebibliography
\def\thebibliography{\DeclareRobustCommand{\VAN}[3]{##3}\VANthebibliography}

\usepackage{graphicx}	
\usepackage{amsmath}	

\title[Early TeV of GRB 221009A]{What absorbs the early TeV photons of GRB 221009A?}

\author[Shen et al.]{
Jun-Yi Shen, $^{1}$
Yuan-Chuan Zou,$^{1}$\thanks{E-mail: zouyc@hust.edu.cn (Y-CZ)}
A. M. Chen, $^{2}$
Duan-Yuan Gao $^{1}$
\\
$^{1}$Department of Astronomy, School of Physics, Huazhong University of Science and Technology, Wuhan, 430074, China\\
$^{2}$Tsung-Dao Lee Institute, Shanghai Jiao Tong University, Shanghai 201210, China \\
}

\date{Accepted XXX. Received YYY; in original form ZZZ}

\pubyear{2023}

\begin{document}
\label{firstpage}
\pagerange{\pageref{firstpage}--\pageref{lastpage}}
\maketitle

\begin{abstract}
The tera-electronvolt (TeV) light curve of gamma-ray burst (GRB) 221009A {exhibits} an unprecedentedly rapid rise at  {its start}.
This phenomenon could be due to the strong absorption of photons and electrons within the emitting region. 
As the external shock expands outward and the radius increases, the volume of matter also  {grows},  {resulting in a gradual reduction of} the optical depth for TeV photons.
 {We investigate several potential explanations for the early TeV light curves.}
We calculate the optical depth for TeV photons,  {considering both} annihilation with lower-energy photons in the external shock and  {their} scattering by electrons  {generated through the} cascading of the TeV emission. 
Even under the  {favorable} assumptions,  {we have determined that} the optical depths for these processes are orders of magnitude too small to explain the observed light curve.  {Additional} sources of absorbers,  {including} electrons in the ejecta or external shock, also do not  {result in} sufficient optical depths. 
 {Hence, the cause behind} the early peculiar TeV light curve remains  {unclear}.

\end{abstract}

\begin{keywords}
(transients:) gamma-ray bursts -- opacity 
\end{keywords}



\section{Introduction}
Gamma-ray bursts (GRBs) are transient sources  {caused by} highly energetic astrophysical events that emit  {copious} number of high-energy photons, which travel  { extremely long} distances to reach the observer.  {The emission in GRBs} can be divided into two  {stages}: the prompt emission and the afterglow emission, each exhibiting distinct observational characteristics. In recent years,  several GRBs have been detected at very high energies (VHE, i.e., $E>0.1\ {\rm TeV}$), including GRBs 180720B \citep{2019Natur.575..464A}, 
190114C \citep{2019Natur.575..455M}, 
190829A \citep{2021Sci...372.1081H},
201015A \citep{2020GCN.28659....1B}, and 
201216C \citep{2020GCN.29075....1B}. 

On October 9th 2022 at 13:16:59.99 UT ($T_0$), the Gamma-Ray Burst Monitor (GBM) onboard the Fermi satellite
triggered on GRB 221009A \citep{2023ApJ...952L..42L}. 
Following the trigger, other high-energy detectors such as Konus-WIND, SRG/ART-XC, INTEGRAL, Insight-HXMT, GECAM-C, Swift, MAXI, and NICER also detected the prompt emission and  {the early} afterglow emission \citep{article1,aa28032023,2023arXiv230301203A,2023ApJ...946L..24W}. With a redshift $z$ $\sim$ 0.151, corresponding to a luminosity distance $ {D_{\rm{L}} \sim }  2.1 \times 10^{27} {\rm cm}$ \citep{2023arXiv230207891M}, GRB 221009A stands out as an exceptionally high luminosity event.  {In fact, it} is the brightest burst detected  {so far} \citep{2023ApJ...946L..31B}.



The Large High Altitude Air Shower Observatory (LHAASO), situated in Daocheng, Sichuan Province, China \citep{2019arXiv190502773C}, also reported the detection of the very early VHE afterglow of GRB 221009A, with more than 64,000 photons above 0.2 TeV observed within the first 3000 seconds \citep{2023Sci...380.1390.}. 
Overall, as shown in Fig. 3 of \cite{2023Sci...380.1390.}, the TeV light curve of GRB 221009A shows a four-segment  {shape: a rapid initial rise}, a slower rise up to the peak, a slow decay after the peak, and then a steep decay. Each stage can be well-fitted by a power-law function of time ($f_{\nu} \propto t^{\alpha}$, where $t=T-T_{\star}$, and $T_{\star}$ is defined as $T_{\star}=T_0+226\ {\rm s}$), indicating an external shock origin \citep{2023Sci...380.1390.}. 
Although the observed TeV data can be  { in general} explained by the synchrotron self-Compton emission in the external shock, the rapid rise at the very early stage ($t\sim \mathbf{1} -4.85\ {\rm s}$) with a temporal slope of $\alpha \approx 14.9$ is difficult to  {explain} under the standard afterglow scenario, which predicts $\alpha = 4$ in a homogeneous medium or $\alpha = 1/2$ in a wind environment \citep[see Eq. (S13) of][]{2023Sci...380.1390.} 



 {The rapid rise of TeV flux could be caused by several processes. First, it could be due to the dynamical process of the external shock. At the very early stage, the external shock -- driven by the outer ejecta -- could be energized by the inner ejecta, leading to the increase of the bulk Lorentz factor, and therefore increasing the TeV flux dramatically \citep{2023Sci...380.1390.}. 
Second, the rapid rise behavior could be due to the strong absorption of photons and electrons within the emitting region. In particular, the TeV flux around $t\sim 1.8\ {\rm s}$ is significantly higher than the average flux during $t\sim 0-4.85\ {\rm s}$, and it can be extrapolated back from the normal slower rise during $t\sim 4.85\ {\rm s}$ to $t\sim18\ {\rm s}$ \citep{2023Sci...380.1390.}.}
This indicates the fast rise phase during $t \sim 2\ {\rm s}$ to $t \sim 4.85\ {\rm s}$ could be due to the  {severe} absorption of the TeV photons within the shock.
In this paper, we will explore the possibility of this  {second} scenario for the physical origin of the early stage of the TeV light curve of GRB 221009A.


 {Initially, in this scenario the external shock is optically thick to TeV photons, but as time progresses, the optical depth ($\tau$) decreases, and the external shock becomes transparent for TeV photons. At the same time, TeV photons could also be blocked by cascade-generated secondary electrons. As the external shock moves outward, the electron density decreases, leading to a rapid increase of the TeV flux at the early stage.} We also explore other potential sources of particles that may absorb TeV photons, such as annihilation with  {keV photons generated by the shock},  {and scattering by the particles within the ejecta}.
%


This paper is organized as follows:  {in} section \ref{sec:trap},  {we} introduce the method of calculating the external shock particles density  {generated by the cascading process}.  {In section \ref{sec:trap}}, we also describe other  {sources of electrons} that may absorb TeV photons. Then, we  {use} the observational data  {of GRB 221009A in our calculation} of the optical depth. In section \ref{sec3}, we present  {our discussion and conclusions.}

\section{Possible scenarios for absorbing the early TeV photons} 
\label{sec:trap}
A TeV photon can  {undergo} two processes:  {1. it can be} annihilated  {by interacting} with a low-energy photon  {(pair production)} and  {2. it can be} scattered by an electron  {(Compton scattering)}.
 {For the first ``annihilation'' process, since the number of TeV photons is significantly lower than low-energy photons, the number deduction of low-energy photons in the annihilation reaction can be ignored.} We can constrain  {the flux of these low-energy photons} directly from the observations.  {After calculating the flux of low-energy photons, we can then calculate the optical depth of TeV photons.} 
 {For the second ``scattering'' process}, the electrons (and/or positrons) may have diverse origins.
The cascade process  {of TeV photons} would generate a large number of electron-positron pairs in the external shock. Additionally, the external shock  {sweeps up} the interstellar medium (ISM) and  {entrains} these particles, causing them to move together with the external shock. 
 {This group of electrons may also block TeV photons.}
 {\cite{2023SCPMA..6689511W} suggested that TeV photons originate from internal shocks, and then, electrons within the internal shock may also absorb TeV photons under this scenario}. 
 {Also, electrons within the ejecta that come from the central engine may also scatter the TeV photons.}
In this section, we will describe how to calculate the optical depth of TeV photons in  {all of these different scenarios}.  

Before  {delving into the origin} of low-energy photons or electrons, we first  {present general arguments in calculating the optical depth of TeV photons. As mentioned above, TeV photons can interact by several processes such as $\gamma \gamma$ pair production and Compton scattering}. 
The  {Compton} scattering cross-section, $\sigma_{\rm c}$,  {can be expressed using the} Klein-Nishina equation \citep[Section 2, Chapter 5 in][]{1998,1929ZPhy...52..853K}: 
\begin{equation} \label{eq:sigmaC}
    \sigma_{\rm c}=\frac{3\sigma_{\rm{T}}}{8}{\epsilon}^{-1}(\ln{2{\epsilon}}+\frac{1}{2}),
\end{equation}
where ${\epsilon}=\hbar \omega /m_{\mathrm{e}}c^2$ is the photon energy in units of $m_{\mathrm{e}} c^2$, and $\sigma_{\rm{T}}$ is Thomson scattering cross-section.



The $\gamma \gamma$ process cross-section is described by the following equation in the head-on collision approximation \citep[Section 7, Chapter 5 in][]{1998,1967PhRv..155.1404G}:
\begin{equation}
 \sigma_{\gamma \gamma} = \frac{3}{16} \sigma_{\rm{T}} (1- \beta ^2) \left[(2- \beta^2)\ln \frac{1+ \beta}{1-\beta}-2\beta (2-\beta^2)\right],
 \label{sgg}
\end{equation}
with 
\begin{equation}
    {\beta = \sqrt{1-\frac{m_{\mathrm{e}} c^2}{ \hbar \omega_0}\cdot\frac{m_{\mathrm{e}} c^2}{ \hbar \omega_0'}},}
\end{equation}
 {where $\omega_0$ and $\omega_0'$ are the frequency of the two photons in the laboratory frame.}

The optical depth $\tau$ of TeV photons can be estimated as
\begin{equation}
    \label{tau}
    \tau\simeq \sigma n l,
\end{equation}
where the $\sigma $ is the cross-section, either from Eq. (\ref{eq:sigmaC}) or Eq. (\ref{sgg}), depending on the  {specific process we will consider below}, $n$ is the number density of the reaction particles and $l$ is the thickness of the region.
From Eq. (\ref{tau}), we can calculate whether  {the TeV} photon can escape from the region.
 {When $\tau \ll 1$, the plasma becomes transparent to TeV photons.}
In the following, we  {will} consider the optical depth for the TeV photons  {scattering with various sources of electrons and the annihilation due to pair production}.

\subsection{Absorbed by external shock photons}
\label{sec2.4}
In this section,  {we discuss the absorption of TeV photons by low-energy keV photons} from the external shock itself. 
 {These low-energy photons are also produced by the afterglow.}
 {Quantum electrodynamics (QED) calculations indicate that}  {$\sigma_{\gamma \gamma} $ is approximately equal to $ \sigma_{\rm{T}} $} and  {reaches its maximum in a head-on collision and  {when} $\hbar \omega_0 \hbar \omega_0'=(m_{\mathrm{e}} c^2)^2$} \citep[refer to Section 7, Chapter 5 in][]{1998}.
In the co-moving frame, the frequencies of the two head-on photons will be reduced to  { $ \nu/[(1+z)\Gamma_{\rm{b}}] $ and $\nu' /[(1+z) \Gamma_{\rm{b}}]$, where $ \nu $ and $\nu' $ are the frequency of the two photons detected.}
Taking $\Gamma_{\rm{b}} = 560$ \citep{2023Sci...380.1390.} for the  {interaction}  {involving} $\sim 1$ TeV photons  {at the condition of } largest cross-section condition,  the lower energetic photon  {has energy} $ h\nu' \sim 0.08$ MeV, which is  {falls within} the X-ray band.

 {Based on the discussion above, we can determine the energy band that interacts with TeV photons in the $\gamma \gamma$ pair production.} We can then calculate the absorption of TeV photons by the external shock keV photons according to the  $\gamma \gamma$ reaction channel. The $\tau$ can be 
 approximately calculated as:
\begin{equation}
    {\tau(t) \approx \frac{D_{ { \rm{L} } }^2}{R^2 (1+z)}\int_0^t\int_{\nu_{\min}}^{\nu_{\max}} \frac{f_\nu (t') \sigma_{\gamma \gamma}}{ h \nu} {\rm d} {\nu} {\rm d}t'}.
\end{equation}
 {The shock radius $R$ is given by:}
\begin{equation}
    R=2 c \Gamma_{\rm{b}}^2 {t/(1+z)}.
\end{equation}
 {Notice the bulk Lorentz factor $\Gamma_{\rm{b}}$ is evolving with time. However, since we are considering the very early stage, that is, $t \sim 1 - 4.85 \rm{s}$, it is before the deceleration time, which shows as a peak at $t \sim 18 \rm{s}$. Therefore, we consider $\Gamma_{\rm{b}}$ to be constant in this work for estimates of order of magnitude. The evolution might be considered in the detailed modeling.}

\cite{2023ApJ...948L..12K}  {presents} X-ray afterglow observations  {of GRB 221009A}, which can be seen in  {their} Fig. 12. 
 {Considering $\alpha \sim -1.3$ and $\beta \sim -0.75$ for  $f_\nu (t) \propto t^{\alpha} \nu^{\beta}$ \citep{2023ApJ...948L..12K}, we can calculate the observed flux density at $t=4$ s and at $\sim 0.08$ MeV (see above), which is about $10^{-25}\ \rm{erg \ cm^{-2} \ s^{-1} \ Hz^{-1}} $.}
 {We now select an} energy range of 0.2 - 2000 keV, as this band yields a large cross-section.  {Even in these favorable conditions, we find an optical depth} 
$ \tau(t)|_{t=4 \, s} \sim 10^{-5} (t/4 {\rm s})^{-0.3}$.
 {This optical depth is too small and its decay rate is too slow. Moreover, it cannot explain the start of the TeV emission.}

\subsection{{Absorbed by the cascaded pairs}} \label{section cascade}
A promising and self-consistent scenario could be  {that} the TeV  {photons are blocked by the}  {cascade-generated secondary} electron- {positron}.
 {We describe the scenario as follows. During}
 the very early time ($ \mathbf{t} < \mathbf{2}$ s), the cascade  {had} not fully  {initiated},  { allowing TeV photons to escape. } Later on ($ \mathbf{t}$ $\sim 2-4.85$ s), the  {effect of the cascade appeared}, and the cascaded electron- {positron} pairs further blocked the TeV photons via the Compton scattering process. This is  {why} we observed the dip in the TeV light curve. After that, with the  {increase of the} radius of the  {external shock}, the optical depth drops to less than unity, and the TeV light curve returned to a  {regular} GRB afterglow in the optically thin case.
The cascade process can be studied using Monte Carlo simulations \citep{2000CoPhC.124..290M}. 
\cite{2013ApJ...768...54B} developed a semi-analytical method to  {simplify the calculation of the number} of cascade particles,  {which we follow below}. 

We roughly consider that only TeV photons  { contribute to system energy injection}.  The observable photons $\dot{N}_{\epsilon}^{\rm esc}$  can be written as follows \citep{2013ApJ...768...54B}:
\begin{equation}
    \begin{split}
        \dot{N}_{\epsilon}^{\rm esc}=(\dot{N}_{\epsilon}^{0}+\dot{N}_{\epsilon}^{\rm sec})\left[\frac{1-e^{-\tau_{\gamma\gamma}(\epsilon)}}{\tau_{\gamma\gamma}(\epsilon)}\right],
    \end{split}\label{11}
\end{equation}
 {where $\tau_{\gamma\gamma}(\epsilon)$ is the optical depth of photons due to $\gamma\gamma$ absorption}, $\dot{N}_{\epsilon}^{0}$ represents the injection rate of TeV photons, and $\dot{N}_{\epsilon}^{\rm sec}$ is the secondary photon component mainly arising from synchrotron radiation,  {which can be expressed as \citep{2013ApJ...768...54B}}:
\begin{equation}
    {\dot{N}_{\epsilon}^{\rm sec}=A_0\epsilon^{-2/3}\int_1^{\infty} {\rm{d}} \gamma_{{\rm{e}}} N_{\rm{e}}( \gamma_{{\rm{e}}} ) \gamma_{{\rm{e}}}^{-2/3} e^{-\epsilon/(b \gamma_{{\rm{e}}}^2)}},
    \label{eq:Nsec}
\end{equation}
 {where $\gamma_{{\rm{e}}}$ is the Lorentz factor of electrons, $N_{{\rm{e}}}(\gamma_{{\rm{e}}})$ denotes the distribution of electrons, $b \equiv B/B_{c},$ where $B_c=4.4 \times 10^{13} $ G and $B$ is the magnetic field strength, and the normalization is given by \citep{2013ApJ...768...54B}:
\begin{equation}
    {A_0=\frac{c \sigma_{\rm{T}}B^2}{6 \pi m_{\mathrm{e}} c^2 \Gamma_{\rm{b}}(4/3) b^{4/3}}}.
\end{equation}
 }
By utilizing Eq. (\ref{sgg}) and Eq. (\ref{11}) as well as the expression for $\dot{N}_{\epsilon}^{\rm sec}$  {(Eq. (\ref{eq:Nsec}))},  {the generation rate of electron-positron pairs $\dot N_{\mathrm{e}}^{\gamma\gamma}$ can be calculated as \citep[see Eq. (10) of][]{2013ApJ...768...54B}: 
\begin{equation}
    {\frac{\partial}{\partial \gamma_{\mathrm{e}}} (\dot\gamma_{\mathrm{e}} N_{\mathrm{e}} (\gamma_{\mathrm{e}})) = \dot N_{\mathrm{e}}^{\gamma\gamma}(\gamma_{\mathrm{e}})+\dot N_{\mathrm{e}}(\gamma_{\mathrm{e}})^{{\mathrm{esc}}}},
\end{equation}}
 {where the $\dot \gamma_{\mathrm{e}}$ denotes the cooling of electrons according to synchrotron radiation, and the $\dot N_{\mathrm{e}}(\gamma_{\mathrm{e}})^{\mathrm{esc}} = -N_{\mathrm{e}}(\gamma_{\mathrm{e}})/t_{\mathrm{esc}}$. The $t_{\mathrm{esc}}$ is the electron escape time scale, expressed as $\eta_{\mathrm{esc}} R/c$, where $\eta_{\mathrm{esc}} \geq 1$. 
These equations give the evolution of $N_{\mathrm{e}}(\gamma_{\mathrm{e}}).$ }

Before  {presenting} the detailed  {semi-analytical} calculation, we can  {make} two estimations for the number of cascaded pairs, i.e., a conservative estimation and an aggressive estimation, to get the lower and the upper bounds. 
 {In} the $\gamma \gamma$   {process}, energy conservation  {dictates that the total energy} of two $\gamma$ photons must be larger than $2 m_{\mathrm{e}} c^2$. Since the external shock is moving with a bulk Lorentz factor $\Gamma_{\rm{b}}\sim 560$ \citep{2023Sci...380.1390.}, the TeV photons are detected in the observer's frame. 
 {Because of the redshift, a factor of $[(1+z)\Gamma_{\rm{b}}]^{-1}$ compared to its energy detected on Earth (see above) should be considered.}
 Therefore, in the observer's frame, the total energy of two photons in the $\gamma \gamma$ process must  {exceed} $E_{ {  \rm{m} } } \simeq 2 m_{\mathrm{e}} c^2 \Gamma_{\rm{b}}(1+z) \sim 660$ MeV. 
 {We assume, as an approximation, that all the energy of TeV photons is converted into the rest mass and kinetic energy of electron-positron pairs. In this condition, the number of cascade pairs is the largest. }
A conservative estimate is that all particles with energy less than $E_{ {  \rm{m} } }$ cannot continue the cascade reaction. Then, we can approximate that TeV photons will cascade and generate $N$ particles, given by:
\begin{equation}
    N=\frac{4 \pi D_{\rm L}^2 \int_{0}^{t} f'(t') \mathrm{d} {t'}}{E_{ {  \rm{m} } }{(1+z)}},
\end{equation}
 {where the flux absorbed by photons, $f'(t')$ can be obtained from observations.} 
By extending the flux  {of TeV photons from 5 }  $\sim$ 10 s to 1 $\sim$ 4.85 s and   {taking the} difference (data are from \citealp{2023Sci...380.1390.}), we can estimate the flux ${f'(t')}$ of  {absorbed} TeV photons in the $\gamma \gamma$ process. 
Then we  {can calculate} the evolution of the optical depth: 
\begin{equation}\label{22}
   \tau(t) {\approx} \frac{ D_{ { \rm{L} } }^2 \int_{0}^{t} f'(t') \mathrm{d}{t'}}{E_{ {  \rm{m} } }  R^2 {(1+z)}} \sigma_{\rm c} .
\end{equation}
For GRB 221009A, with a luminosity distance of $D_{\rm L}=2.1\times 10^{27}$ cm  {we find} a flux $f'(t) \lesssim 10^{-6} \ \rm{erg \  cm^{-2} \ s^{-1}}$.  {We simplified the integral $\int_0^t f'(t') \rm{d} t' $ as $f_{\max}'(t) t$, where $f_{\max}'(t) $ is the maximum of $f'(t')$. With this simplification, we can estimate the upper limit of $\tau$. }  {With Eq. (\ref{eq:sigmaC}) and Eq. (\ref{22}), $\tau(t)$ can be written as}:
\begin{equation}\label{cons tau}
     \tau(t) \sim 1 \times 10^{-6} \frac{f'_{-6} D_{{\rm{L}},27}^2   }{E_{{\rm{m}},2}  \Gamma_{{\rm{b}},2}^4{\epsilon}_3 t}  \sim 2 \times 10^{-9} t^{-1},
\end{equation}
where $Q_{i}=Q\times 10^{i}$, and $E_{ {  \rm{m} } }$ is in unit of MeV,  {other quantities are in unit of cgs units}.   {This calculated optical depth at $t=4.85$ s is significantly lower than 1, indicating that TeV photons are free to leave this area. }
Therefore, the conservative estimation of the optical depth is not sufficient to block TeV photons.

For an aggressive assumption, we choose $E_{m} = $ 0.511 MeV, i.e., assuming the TeV photons are transferred into pairs with no energy waste. 
In this condition, the number density of  {electron-positron pairs} is highest. However, it is impossible for the cascade matter to be so dense,
 {since using} Eq. (\ref{cons tau}), we have:
\begin{equation}\label{eq9}
    \tau(t) \sim 1 \times 10^{-6} t^{-1}.
\end{equation}
We can see that even with the most aggressive estimation, the number density of electrons is not enough to block the TeV photons.

Therefore, without knowing the details of the cascade, we can see that the cascaded electron- {positron} pairs are not dense enough to block the early TeV photons.
 {If we ignore the data at $t \sim $ 1.8 s, the absorption begins at the initial time.}   {We discuss this in the next subsections.}

\subsection{Absorbed by electrons from the external shock}
The TeV photons are believed  {to originate} from the external shock, 
 {where many electrons may scatter the TeV photons.}
We check the optical depth  { for TeV photons} in this case. 
The number density of electrons in the external shock is $\sim \Gamma_{\rm{b}} n_0$, where $n_0$ is the number density of  {the} circum-burst medium. 
The optical depth is $\tau = 2 \sigma_{\rm c} n_0 c \Gamma_{\rm{b}} t$, which is about $10^{-16}$ for typical values. 
 {Moreover,} $\tau$ is proportional to the  {observed} time, which is contrary to the expectation.
Therefore, these electrons cannot absorb the TeV photons.

\subsection{Absorbed by electrons from the ejecta}
\cite{2023SCPMA..6689511W} studied the  {potential for} the prompt phase  {to generate} the TeV photons emission from a hadronic process. \cite{2023ApJ...947L..14Z} considered the possibility that the very high energy photons are from the reverse shock of the external shock. 
In both scenarios, the ejecta from the central engine may radiate the TeV photons. 

Here, we assume  {that the TeV photons are generated in the} internal shock as \cite{2023SCPMA..6689511W} described. We consider the condition of whether the internal shock electrons can block the TeV photons.  {In the internal shock, }the number of electrons $N_{\rm e}$ is: 
 \begin{equation}\label{3}
     N_{\rm e}=\frac{\eta E_{\rm \gamma, iso}}{ \Gamma_{\rm{b}} m_{ {\rm{p}} } c^2 },
 \end{equation}
where $E_{\rm \gamma, iso}$ is the isotropic equivalent energy of the GRB prompt emission, $m_{ {\rm{p}} }$ is the rest mass of the proton, and $\eta$ is the ratio between the kinetic energy  {and} the gamma-ray energy, which is roughly 1.  

Taking $ E_{\rm \gamma, iso} \sim 1\times 10^{55}$ erg and $\Gamma_{\rm{b}} \approx 560$ \citep{2023Sci...380.1390.}, we can calculate the electron number  {in the} internal shock, which is about $2 \times 10^{55}$. 

The optical depth is $\tau= N_{\rm e} \sigma_{\rm c} / [4 \pi (2 c \Gamma_{\rm{b}}^2 t)^2 ]$ {${\approx 3 \times 10^{-4}}$  for ${t\sim 10^{-1}}$ s, which is a conservative choice of the internal shock timescale, and the Lorentz factor $\Gamma_{\rm{b}} \approx 560$ is also a conservative choice for the internal shock.}
Considering the optical depth decreases with time as $t^{-2}$, it becomes even smaller at a later time.
Therefore, in the internal shock scenario, the electrons from the ejecta cannot block the TeV photons. 

For the reverse shock scenario, as suggested by \cite{2023ApJ...947L..14Z}, the condition is the same. The difference is  { that the number of electrons in reverse shock evolves with time, and consequently, the total number of shocked electrons increases, i.e., the thickness $l$ of the shocked region increases.} However, in the calculation of the optical depth, $l$ is canceled. 

\section{Discussion and Conclusions}\label{sec3}
 {We have examined several scenarios to explain the unprecedented rapid rise of the TeV light curve of GRB 221009A, including the absorption and scattering of the TeV photons by the photons or electrons generated within the shock.} 
However, even if all the TeV photons' energy is converted to the rest mass of electrons, it is still not dense enough to block the TeV photons. 
Additionally, we attempted to use external shock-accelerated electrons to block the TeV photons, but it was not successful either. 
We also checked using the afterglow itself to explain this phenomenon, but the flux of afterglow in the hundred keV photons is not high enough to absorb TeV photons. 
In conclusion, the early dip in the TeV light curve still remains to be explained.

There are some other processes that may absorb the TeV emission.
For example, the early TeV photons may also collide with the cosmic microwave background and/or the galactic infrared background. 
However, these sources do not change over time, which is not consistent with the idea that only the early TeV photons cannot escape.

The prompt MeV photons may also absorb the TeV photons. 
Though this is generally considered unlikely as we have seen a full afterglow-like light curve. 
    However, in the early  {times, as the TeV radiation radius can be small, it makes the absorption possible. 
We are not able to estimate the optical depth simply similar to the treatment in section \ref{sec2.4}, as the MeV photons are believed {to originate} from internal shocks, which is different from the TeV}   {origin}.
For this possibility, one should consider carefully the geometry configuration of the MeV-TeV collision, as well as the time and spectral evolution of the prompt MeV emission.

It could be that there was no absorption at all if we neglect the very first emission at around 1 s. 
 {The} early light curve before 5 s can be taken as a fast rise. Such a fast rise might be explained by energy injection to the external shock,  {which has also been suggested in \cite{2023arXiv230900673K} very recently}. If this is the case, one should consider what kind of energy injection can power  {such a} fast rise.


\section*{Acknowledgements}
We thank  {the critical comments from the anonymous referee,} the helpful discussions with Weihua Lei, Kai Wang, Xiang-Yu Wang,  and the hospitality of Yao'an station of Purple Mountain Observatory.
The English is polished by ChatGPT.  {We are very grateful to Rodolfo Barniol Duran for his meticulous manuscript reading and polishing.}
This work is supported by the National Key R\&D Program of China (2022SKA0130100) and China Postdoctoral Science Foundation (2023T160410).

\section*{Data Availability}
No data was used for the research described in the article.




\bibliographystyle{mnras}
\bibliography{main} 





\bsp	
\label{lastpage}
\end{document}